# *System identification calorimetry*

B. P. MacLeod*,[1,4] (ORCID: 0000-0002-8547-9318)
D. K. Fork*,[2] (ORCID: 0000-0001-9559-1277)
B. Lam[1]
C. P. Berlinguette*,[1,3,4,a] (ORCID: 0000-0001-6875-849X)

[1]*Department of Chemistry, University of British Columbia, Vancouver, V6T1Z1, Canada*
[2]*Google LLC, 1600 Amphitheatre Pkwy, Mountain View, California, 94043, USA*
[3]*Department of Chemical and Biological Engineering, University of British Columbia, Vancouver, V6T 1Z3, Canada*
[4]*Stewart Blusson Quantum Matter Institute, University of British Columbia, Vancouver, V6T 1Z4, Canada*
a) Author to whom correspondence should be addressed. Electronic mail: cberling@chem.ubc.ca

## Abstract

We report a model-based method for quantifying heat flow and storage in thermal systems using data from multiple thermal sensors. This approach avoids stringent requirements on the system geometry and sensor positions and enables calorimetry to be performed under a broader range of circumstances than is accessible with existing calorimeters, such as when non-linear heat transfer occurs, when spatially separated heat sources are active, or when multiple thermal masses participate. Using experimental data from a model thermal system, this paper provides a tutorial on the construction of non-linear lumped element heat transfer models and the use of system identification to estimate the parameters of these models from calibration data. The calibrated models are then used to estimate unknown energy inputs to the thermal system from sensor data. Our best model enabled the measurement of the total input energy with 0.02% accuracy; the instantaneous input power could be measured with a root-mean-square error of 10% of the average input power.

## Introduction

Calorimetry provides useful information on the amount and rate of heat evolution or absorption from physical and chemical transformations, electronic devices, batteries and other systems. However, not every application is amenable to the use of established calorimetry techniques, such as when a large or complex device not readily integrated into a calorimeter is the object of study. For complex thermal systems such as buildings [1] and rapid thermal annealers, [2] the thermal response at specific locations to heat inputs can be predicted by a model of substantially lower order than a full finite element model. Here we show that low-order models can also be used to perform calorimetry on a complex thermal system by computing estimates of heat inputs from thermal sensor data.

Many useful calorimeter models and calibration methods have been studied and applied [3]. Lumped-element models have long been employed to analyze the differential scanning calorimeter (DSC), [4] and a number of such models of increasing sophistication are described by Hohne et al [5]. For calorimetric analysis of a system of *n* isothermal bodies undergoing linear heat exchange, Zielenkiewicz [6] has outlined the governing system of equations. However, not all calorimeters are adequately described by linear models. For example, Drebushchak identified that linear models are inadequate for a full description of the DSC, and developed a non-linear temperature dependent equation for the DSC calibration coefficient [7].

We originally developed the method described herein to account for heat in a high-temperature, high-pressure non-differential calorimeter [8] in which the non-linear heat transfer, non-ideal geometry and challenging operating conditions required more sophisticated models than were available. We refer to this method as system identification calorimetry. We contend that many factors may require more complex thermal models including:

1. Non-linear heat transport (e.g., by radiation).
2. Power input that may be distributed spatially (e.g., that provided by a heater winding).
3. Power input that may enter the system from multiple physically separate sources such as sample chambers, heater elements, stirring devices, or catalyst beds.

4. The need to situate temperature sensors away from heat sources that could damage the sensors.
5. Systems with multiple thermal sensors that require sensor fusion models to accurately interpret the response of the system.

The system identification calorimetry method used in this report employs non-linear, lumped element thermal models to describe heat transfer within a thermal system. Once a model topology has been formulated, estimates of the optimal model parameters are obtained numerically from a calibration data set. This type of modelling, in which empirical parameter estimates are obtained for approximate models using calibration data, is known as grey-box system identification [9,10]. Our system identification approach to calorimetry allows us to develop models which can adequately capture heat flows in thermal systems which may be impossible to integrate into conventional calorimeters.

System identification has long been applied to the analysis of data from conventional calorimeters for the purpose of deconvoluting the calorimeter response from the true rate of heat evolution by a sample [11–13]. More recently, Jesus et al. [14] used grey-box system identification to obtain physical insights into the operation of a commercial isothermal titration calorimeter; they determined that at sufficiently high titrant injection rates the calorimeter sensitivity decreased because the region of reagent mixing moved partially outside of the calorimeter. Despite the utility of system identification for calorimetric data analysis, the use of system identification to enable calorimetry in non-standard experimental conditions has been explored in only a single previous body of work, which examined calorimetry on electromagnetically levitated molten metal samples [15–17]. Non-linear grey box system identification has been applied to modelling of heating processes [18,19] but not - with the exception of our preliminary work [8] - to the modelling of thermal systems for the purpose of extracting calorimetric information.

Here, we offer a tutorial-style introduction to the application of nonlinear grey box system identification to calorimetry. We introduce equivalent circuit thermal models then describe the use of system identification to estimate the equivalent circuit thermal model parameters from experimental calibration data. We then provide a general protocol for performing calibrations and calorimetric measurements on thermal systems using our method. Methods for analyzing the power and energy resolution of the resulting calorimetric measurements are introduced. Finally, we provide three worked examples using experimental data from a model thermal system which has linear and non-linear thermal conductances and is instrumented with multiple thermal sensors. These examples show that lumped element models can be used to obtain calorimetric information by combining data from multiple thermal sensors. Our best model, a non-linear lumped element model with only three states, enabled the measurement of the total input power to the system with a 10% root-mean-square error and accounted for the total energy input to the system to 0.02% accuracy.

## Equivalent circuit thermal models

Lumped element thermal models are analogous to $RC$ electrical circuits, in that they obey the same differential equations [20]. Table 1 shows the equivalent time varying quantities and parameters. We illustrate these equivalents in a circuit diagram, shown in Fig. 1, which labels each element and node with both electrical and thermal symbols.

Table 1 Electrical and thermal equivalents

| Electrical parameter | Symbol | Units | Thermal Equivalent | Symbol | Units |
|---|---|---|---|---|---|
| Voltage | $v$ | Volts | Temperature | $T$ | K |
| Current | $i$ | Amperes | Heat flow | $Q$ | Watts |
| Resistance | $R$ | Ohms | Conductance | $k$ | Watts/K |
| Capacitance | $C$ | Farads | Heat Capacity | $c$ | Joules/K |

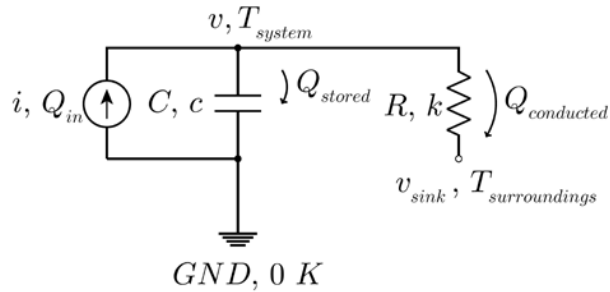

**Fig. 1** Equivalent circuit model of a simple thermal system heated to temperature $T_{system}$ by an input power $Q_{in}$. Heat is stored ($Q_{stored}$) in the thermal mass of the system ($c$) and heat is conducted ($Q_{conducted}$) through the thermal conductance of the system ($k$) to the surrounding environment, which is at a temperature $T_{surroundings}$

Lumped element thermal models approximate the flow of heat through volumes by temperature differences across discrete conductance elements such as the one depicted between temperatures $T_{system}$ and $T_{surroundings}$ in Fig. 1. Volumetric thermal energy storage is approximated by discrete thermal masses which store energy at a single temperature. Under appropriate conditions, these approximations can be remarkably good. The temperature surrounding a thermal system may not be spatially uniform as Fig. 1 suggests by its single conductive path to $T_{surroundings}$. Best practice calorimeter design attempts to create a uniform thermal surrounding or isoperibolic boundary. When this is not possible, it may be necessary to include multiple thermal sinks to obtain an adequate model.

Not all electrical circuits have thermal equivalents. For example, our circuit diagram illustrates the important detail that unlike capacitors in electrical circuits, which can be flexibly placed between any pair of voltage nodes, heat capacities may only connect a temperature node representing a thermal mass, to an absolutely stiff thermal ground, which may be thought of as being at 0 K. This requirement follows from the essential physics that heat capacity captures the thermal energy stored in a mass heated from absolute zero. Now that the electrical analogy is illustrated, in what follows, we will only use the thermal symbols shown in Fig. 1 and Table 1.

## Estimation of thermal model parameters using system identification

Here, our use for system identification is to estimate the set of thermal model parameter values $\theta$ that best explain past, present and future thermal sensor measurements given known inputs. In the example above, the measured temperature is $T_{system}$, the inputs are $Q_{in}$ and $T_{surroundings}$ and $\boldsymbol{\theta} = \{c, k\}$. Note that neither $c$ nor $k$ nor other parameters in general are actual *measurements* of thermal masses or conductances of clearly identified parts of the physical apparatus; but they are related in the sense that they are weighted averages of the many small thermal mass elements and conductive paths that describe the real spatially distributed thermal system. This notion illustrates the importance of choosing carefully where the state of the system, $T_{system}$, is measured because that will often determine model quality.

The differential equation corresponding to the circuit in Fig. 1 is shown below

$$\frac{dT_{system}}{dt} = \frac{1}{c}[-k(T_{system} - T_{surroundings}) + Q_{in}] \qquad (1)$$

We have written Eq. 1 according to the convention [21] for a state space representation, which for linear systems takes the form

$$\frac{d\boldsymbol{x}}{dt} = \boldsymbol{A}\boldsymbol{x} + \boldsymbol{B}\boldsymbol{u} \qquad (2)$$

where $\boldsymbol{x}$ is the state vector, which in Eq. 1 is simply $T_{system}$, $\boldsymbol{A}$ is the system matrix, $\boldsymbol{u}$ is the input vector, which in Eq. 1 is $(T_{surroundings}, Q)$ and $\boldsymbol{B}$ is the input matrix. For non-linear systems, the equation of state takes the more general form

$$\frac{d\boldsymbol{x}}{dt} = f(\boldsymbol{x}, \boldsymbol{u}) \qquad (3)$$

In addition to the equation of state, the model must have an output, generally denoted by the vector $\boldsymbol{y}$ which represents the quantities that the model seeks to predict. In this example, the output is simply $T_{system}$; however, other outputs may be more practical in some contexts. For example, if temperature is measured with a resistance temperature detector and the calibration is not known, the output $\boldsymbol{y}$ could take the form $R = R_o + \alpha(T_{system} - T_0)$ and the fitting parameters $R_o$, $\alpha$ and $T_o$ become part of the model parameters $\theta$. We note several important points here:

1. Fig. 1 depicts a "grey box model," the only type considered here. As the name suggests, the user may know something about the internal physics of how the model works but seeks to estimate the value of model parameters empirically. This is an important distinction from "black box models" whose structure is not physically motivated, may be non-causal and may not conserve energy.
2. Fig. 1, and all other models written as equivalent circuits, conserve energy. This is important because provided that we carefully represent every power input into the apparatus with corresponding inputs in the model, the calibration will make predictions that conserve energy. Energy conservation is, in this context, a consequence of enforcing Kirchhoff's current law for a thermal equivalent circuit.
3. What we call "calibration" in calorimetry is called "estimation" in system identification. Estimation comprises finding the best parameter values $\widehat{\boldsymbol{\theta}}$, which in the present example are the values of $c$ and $k$, that minimize the model output prediction error,

$$\boldsymbol{\varepsilon}(t, \boldsymbol{\theta}) = \boldsymbol{y}_{measured}(t) - \boldsymbol{y}_{predicted}(t, \boldsymbol{\theta}, \boldsymbol{u}(t)) \qquad (4)$$

where $\boldsymbol{y}_{predicted}(t, \boldsymbol{\theta}, \boldsymbol{u}(t))$ is the time-dependant vector of outputs predicted by the model for the given parameters and inputs. Stated as an optimization problem,

$$\widehat{\boldsymbol{\theta}} = arg\,min_{\boldsymbol{\theta}}|\boldsymbol{\varepsilon}(t, \boldsymbol{\theta})| \qquad (5)$$

where the optimal parameters $\widehat{\boldsymbol{\theta}}$ minimize the L[1] norm of all the prediction errors.

Estimating the parameters can in general be a computationally intensive task, especially if the model is complex or non-linear and the dataset is large. For parameter estimation, we use the System

Identification Toolbox provided by MATLAB® to make this task considerably easier. We will make reference to specific System Identification Toolbox features below.

## Calibration, validation and measurement protocol

Below we outline our protocol for calibrating, validating and making measurements using a calorimeter model. This is easiest when a suitable model starting point, such as the one state model described above, is already known. Adding model features to improve the prediction is typically an iterative process.

*Define the thermal system inputs and outputs*

1. **Determine the bandwidth and amplitude limits of the power inputs, $u_{experiment}(t)$ over which the ability to make calorimetric measurements is desired**.
2. **Determine the measurement time, $t_{meas}$.** This is a practical tradeoff between how long one is willing to spend to gather a feature-rich dataset, and how accurately one wants to determine model parameters. Typically, one must gather calibration data for a multiple of the instrument's thermal relaxation time.
3. **Develop time varying calibration power inputs, $u_{calibration}(t)$, that excite the calorimeter well beyond the bandwidth and amplitude of the power inputs, $u_{experiment}(t)$, to be used during measurement.** Ideally, calibration datasets will also include two additional features:
    a. a period with no input power where all temperature sensors are at the same temperature to measure sensor offsets
    b. multiple periods, with different but constant input powers, where the system is at thermal equilibrium. These periods facilitate the discrimination of thermal conductances (which affect the equilibrium temperatures of the system) from thermal masses (which do not).

*Specify, improve and validate the model*

4. **Specify a model of the calorimeter state responses, $\frac{dx}{dt} = f(x, u)$ and measured outputs $y = g(x, u)$**. Typically, this entails writing a small computer program that produces $\dot{x}$ and $y$ from $\theta$, $x$ and $u$. Using Matlab, grey box models are specified with function calls to either `idgrey` or `idnlgrey`.
5. **Estimate the model parameters**, $\theta$, by fitting the model to the calibration data $(u_{calibration}(t), y_{calibration}(t))$. Using Matlab, grey box models are estimated with calls to either `greyest` or `nlgreyest`.
6. **Examine the model fit and residuals $e_{y_i} = y_{i,measured} - y_{i,predicted}$** to identify any major model failures
7. **Quantify how well the model accounts for power and energy throughout the calibration dataset:**
    a. Derive the modeled heat flows $Q(t) = h(x(t))$ that store energy in the calorimeter, or flow power across the measurement boundary.
    b. Compare the simulated heat flows $Q_{simulated}(t) = h(x_{simulated}(t))$ with the heat flows inferred from the measured data $Q_{inferred}(t) = h(x_{inferred}(t))$. Here $x_{simulated}(t)$ is the state trajectory predicted by the model based on the inputs $u$ and $x_{inferred}(t) = g^{-1}(y_{measured}, u)$ is an estimated state trajectory calculated from the experimental data and the system inputs. This will be explained further below.

    If the power and energy accounting is considered adequate, continue; if not, perform further model development (step 4).

8. *Validate the calibrated model*
    a. Predict the system response $y$ to an input $u(t)$ which is different than $u_{calibration}(t)$ and which respects the bandwidth and amplitude limitations defined in step 1.
    b. Quantify the power and energy accounting for the validation data as was done for the calibration data in step 7b. The reproducibility of the measurement should also be quantified through repeated experiments.

*Conduct thermal measurements*

9. **Conduct calorimetry experiments** with power inputs $u_{experiment}(t)$ which respect the bandwidth and temperature limitations defined in step 3.
10. **Infer the heat flows of interest using the calibrated model**, the experimental inputs $u_{experiment}(t)$ and the experiment data outputs $y_{measured}(t)$.

# Power and energy analysis

Since the purpose of a calorimeter is to analyze thermal power and energy, this section describes their relationship to the equivalent circuit models described above. The thermal system in Fig. 1, and also more complex systems, must conserve energy by storing, or conducting away energy that goes in:

$$Q_{in} = Q_{stored} + Q_{conducted} \tag{6}$$

For the one state model described above, the stored heat flux is what goes into the heat capacity

$$Q_{stored} = c\frac{dT_{system}}{dt} \tag{7}$$

and the conducted heat flux is

$$Q_{conducted} = k(T_{system} - T_{surroundings}) \tag{8}$$

We can gain useful insight to the power resolution of the calorimeter by inspecting the power residuals, the differences between the predicted and measured powers during a control run (no sample). For the one state model described above the conducted power residual, $e_{Q,conducted}$ is simply

$$e_{Q,conducted} = k(T_{measured} - T_{predicted}) \tag{9a}$$

The stored power residual, $e_{Qstored}$ is simply

$$e_{Q,stored} = c(\frac{d}{dt}T_{measured} - \frac{d}{dt}T_{predicted}) \tag{9b}$$

If another source of heat, such as a sample, is added to the calorimeter, and its heat source is not treated as an input to the model, then the sample heat will show up in the power residuals.

Time integral quantities of power

$$U(t) = \int_{t=0}^{t} Q(t')dt' \tag{10}$$

are useful because they measure the energy added, stored, and removed over time. Energy conservation requires these amounts to balance. The energy residuals, during prediction control runs reveal the energy resolution of a calorimeter. Energy resolution, as a fraction of the total energy that has entered the system is often better than power resolution as a fraction of the instantaneous power, especially when the calorimeter conditions are changing rapidly.

# Three worked examples

Here we demonstrate the protocol above with three worked examples selected to further illustrate the following concepts:

- Example 1: three-state linear model
    - Multi-state behaviour
- Example 2: three-state non-linear model
    - Non-linear parameter fitting
- Example 3: three-state non-linear model with multiple sensor types
    - Replacing conventional temperature measurement with native sensor output calibration
    - Hidden state modeling

These examples all use experimental data from a model thermal system constructed in our laboratory. The first two examples use linear and non-linear versions of a three state lumped element heat transfer model of that apparatus to model the response of discrete temperature sensors to heat inputs. The third example uses a three-state model which replaces one of the temperature measurements with the signal from a distributed heat-flow sensor.

## *Description of the apparatus*

The experimental data used here were obtained from a model thermal system (Fig. 2) consisting of a heat source exhibiting non-linear conductance to its environment placed inside a non-differential heat flow calorimeter. The heat source (Fig. 3) consists of a power resistor and two thermometers inside of a sealed plastic bottle partially filled with deionized water. Vaporization and condensation of a liquid can result in intense, highly non-linear heat transport and are exploited in heat pipes for this reason [22]. Here, we exploit vaporization and condensation to introduce a non-linear thermal conductance into our model thermal system. The calorimeter (Fig. 4) is a simplification of a known design [23,24] and consists of an aluminum isothermal enclosure with its interior surfaces tiled with an array of series-wired Seebeck-effect heat-flow sensors (thermopiles).

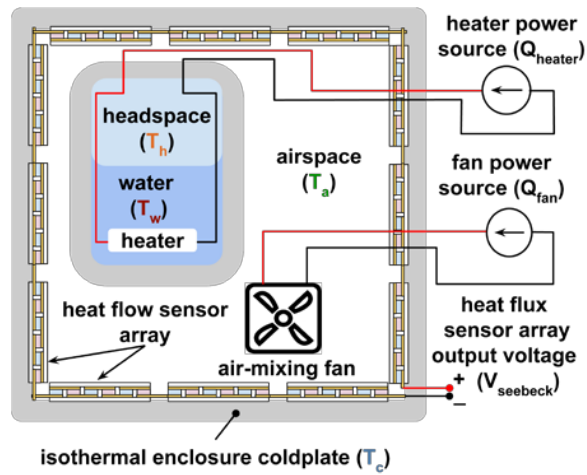

**Fig. 2** Schematic representation of the heat source and calorimeter. The heat source is a sealed plastic bottle containing a heater (with input power $Q_{heater}$) immersed in deionized water. Platinum resistance temperature detectors (RTDs) are provided to measure the temperature of: the water ($T_w$); the headspace above the water inside the bottle ($T_h$); the airspace inside the calorimeter ($T_a$); and, the isothermal enclosure

coldplate ($T_c$) . A fan, with input power $Q_{fan}$, mixes the air in the airspace to promote an even distribution of heat to the heat flow sensor array

The temperature of the isothermal enclosure is regulated by heat exchange fluid that circulates through fluid channels in the enclosure's aluminum walls. A fan circulates air within the enclosure to deliver heat to the inner surface of the Seebeck sensor array, which produces an open circuit voltage $V_s$ linearly proportional to the temperature difference (or equivalently the heat flow) across the array as demonstrated by the calibration data summarized in Fig. S1 of the electronic supplementary material (online resource 1).

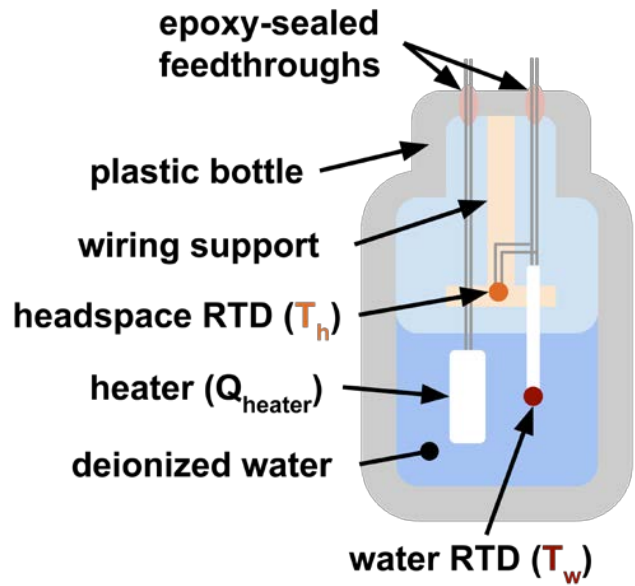

**Fig. 3** Diagram of the heat source. The heater (a 10 ohm power resistor) is placed in a sealed plastic bottle. The bottle is partially filled with ~50 mL of deionized water. Platinum resistance temperature detectors (RTDs) are provided to measure the water temperature and the temperature in the headspace above the liquid. Non-linear thermal conduction from the power resistor to the bottle wall arises due to evaporation at the surface of the liquid followed by condensation elsewhere. The power resistor and water RTD were coated with electrically-insulating epoxy to prevent short circuits through the water in which they are immersed

The experiment is run using custom LabView software to control and acquires data from a variety of digital instruments. The input power to the heater is delivered and measured by a source measure unit (Keithley 2410). The power to the fan was delivered and measured by a programmable power supply (Keithley 2231A-30-3). The resistances of the platinum resistance temperature detectors were measured using a digital multimeter (Keithley 2000) with a multiplexer card (Keithley 2000-Scan). The Seebeck voltage is measured with a nanovoltmeter (Keithley 2182A). Data were acquired at 0.3 Hz. To speed up computations, all results shown here were computed with the data further downsampled by a factor of 5 to an effective sampling frequency of 0.06 Hz (sampling period of 16.66 s)

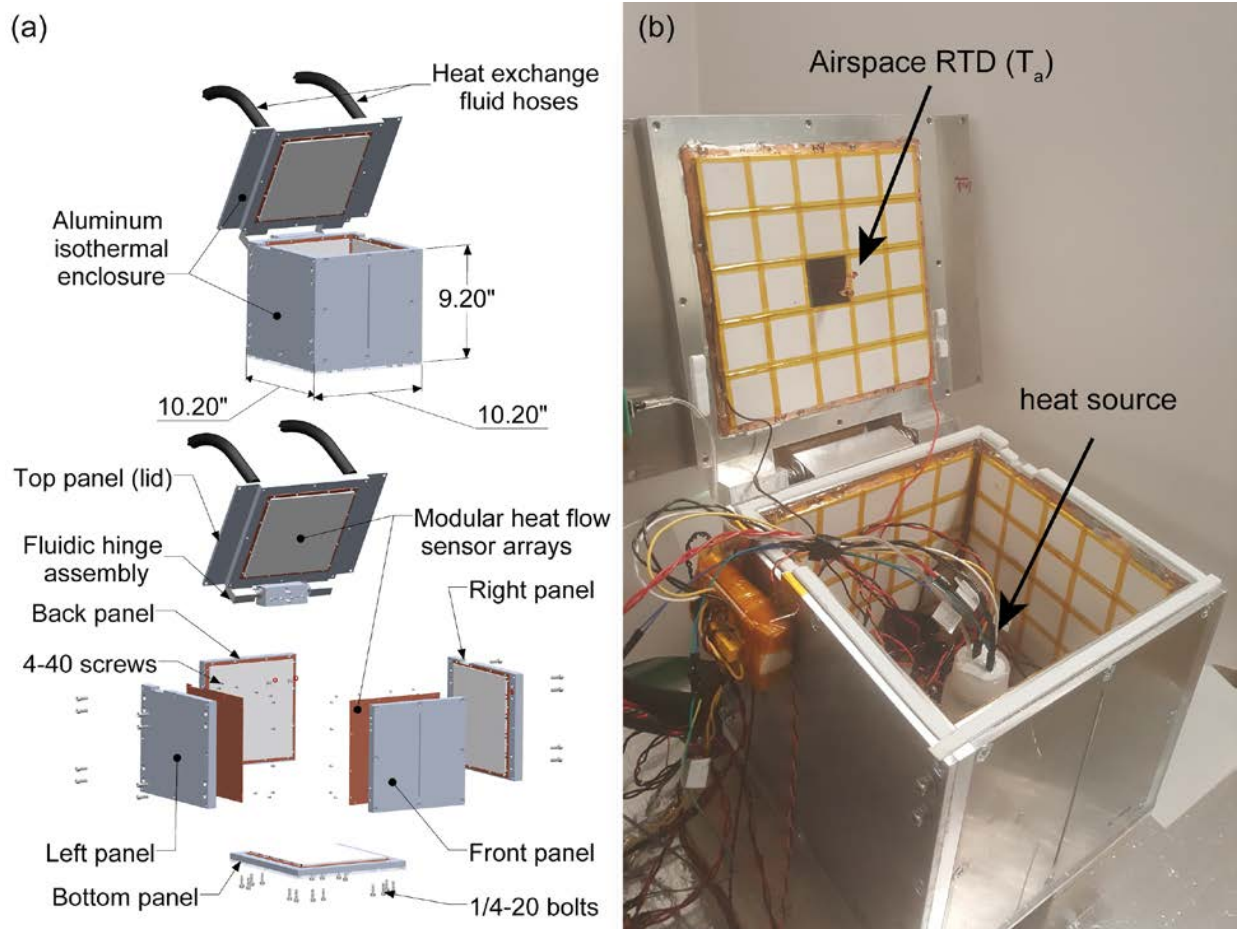

**Fig. 4** Overview of the heat flow calorimeter design. **a** Mechanical diagram of the calorimeter. The calorimeter consists of six modular heat flow sensor arrays anchored to a liquid-thermostatted aluminum isothermal enclosure. The sensor arrays are built by adhering thermoelectric elements (generic, P/N TEC1-12706) to a copper plate and electrically connecting the elements in series. The modular sensor arrays are screwed to the inside of the aluminum isothermal enclosure. The isothermal enclosure consists of six panels, each with an internal serpentine fluid path for heat exchange with a flowing heat transfer fluid. The heat transfer fluid is provided by a recirculating temperature-controlled bath (not shown, VWR, 1187P). The top panel of the isothermal enclosure is hinged to allow access to the inside of the calorimeter; the hinge is a fluidic type which provides a fluid path through the tubular hinge pin. All six panels of the isothermal enclosure are fluidically in series. The fluid flow path runs from the inlet hose into the lid, through the fluidic hinge into the remainder of the enclosure and then returns to the lid via the fluidic hinge before exiting the apparatus via the outlet hose. A small fan (not shown, Digikey P/N 603-2014-ND) is provided to thoroughly mix the air inside the calorimeter, ensuring that heat from the heat source is distributed uniformly to the sensor arrays. **b** Photograph of the calorimeter as constructed, with the lid open. The temperature of the airspace ($T_a$) inside the calorimeter envelope is measured by a platinum RTD affixed to piece of copper foil and suspended from the lid. The heat source is in position in the photograph. The calorimeter is housed in a homebuilt, temperature controlled air bath (not shown) which is stabilized to ± 5mK with a filament heater, platinum RTD and a PID temperature controller. Also not shown is a platinum RTD affixed to the aluminum coldplate of the isothermal enclosure to measure its temperature

### *Description of the experimental procedure and resulting data*

For step 1 of the calibration and measurement protocol, we chose a heater power range of 0 - 4 W (corresponding to a $T_w$ temperature range from ambient up to ~35 °C) as our measurement condition of interest. For step 2, we determined that several days of calibration data would be adequate, given the ~3 hour thermal time-constant of the apparatus. We then performed an 80 hour experiment was performed in which time-varying power was applied to the heater. The experiment included an initial settling time (0 - 3 h), a calibration segment (3 - 40 h) and a validation segment (40 - 80 h). The calibration segment comprised input power pulses with amplitudes up to 5 W (exceeding the 4 W power range to be used for validation/measurement as per protocol step 1). The validation segment, which here doubles as an example measurement, uses lower amplitude and bandwidth inputs as protocol steps 8a and 9. The fan, which dissipates ~160 mW, was powered on for the duration of the experiment. The heater and fan power, the measured temperatures and Seebeck voltage during the experiment are shown in Fig. 5.

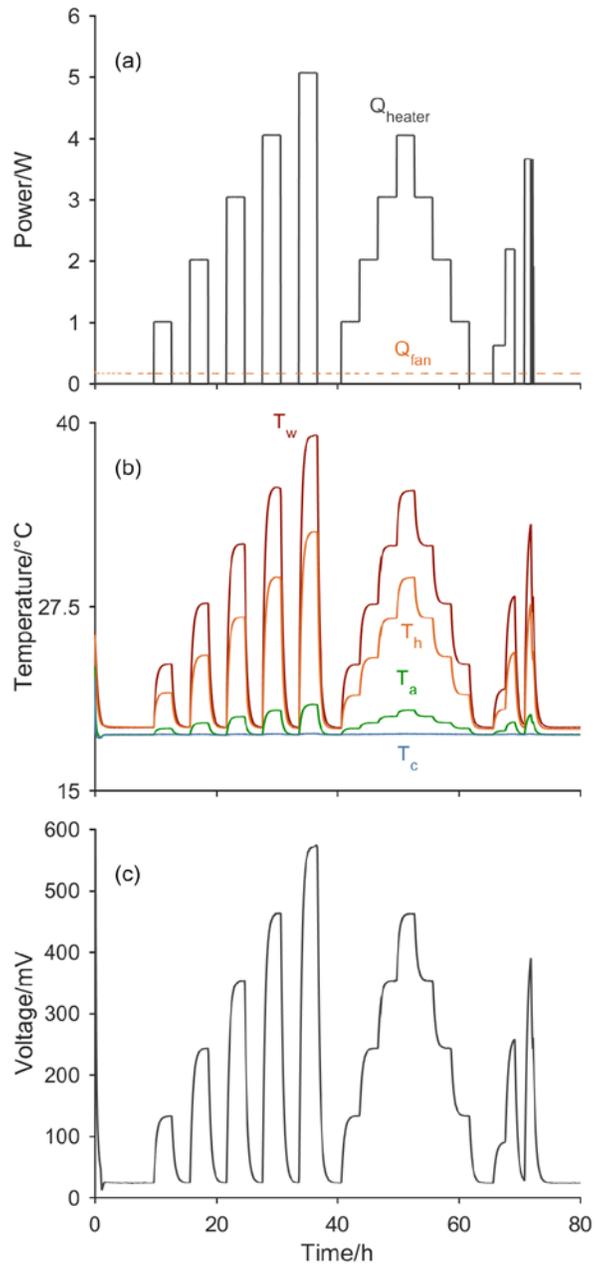

**Fig. 5** Raw data from the experiment. **a** heater and fan power **b** temperatures measured in the heat source water ($T_w$) and headspace ($T_h$), in the calorimeter airspace ($T_a$) and on the calorimeter cold plate ($T_c$) **c** Seebeck voltage from the heat-flow sensor array. The calibration segment is the high-amplitude, high-bandwidth input power pulse train occurring between hours 3-40 (this range excludes initial transient signals). The validation segment is the lower-amplitude set of steps and pulses between hours 40-80

### *A three-state heat transfer model of the apparatus*

The temperature sensor locations in the apparatus were chosen to measure temperatures representative of large areas (e.g the cold-plate) or volumes (e.g. the various fluid spaces) within the apparatus; these temperatures are plausible state variables in a physically descriptive model. Developed the three-state lumped-element heat transfer model of the apparatus (Fig. 6) by placing a thermal mass at each sensor location / temperature-node. The nodes were then connected with thermal conductances in a topology based on the geometry of the apparatus, with the water and headspace nodes transferring heat to the airspace node, and the airspace node in turn transferring heat to the cold-plate. A thermal conductance was also added to connect the headspace and water temperature-nodes.

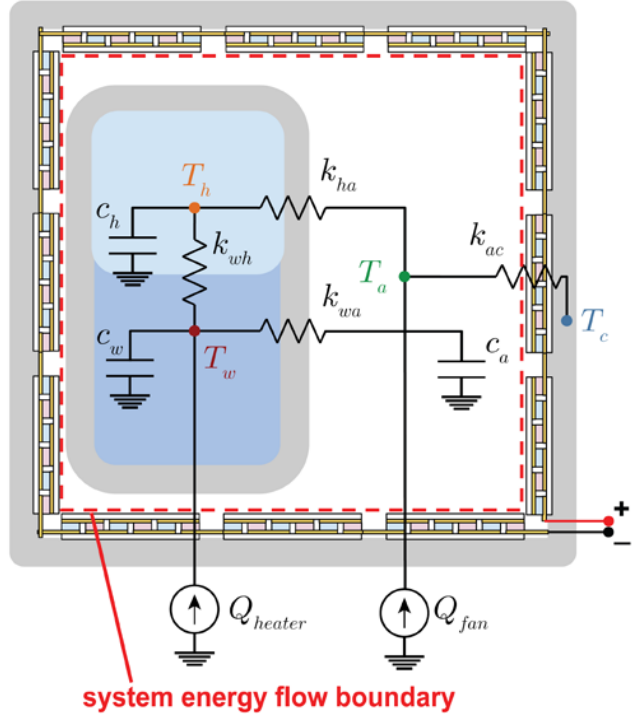

**Fig. 6** A three-state lumped-element heat transfer model of the apparatus. The model is superimposed on the apparatus to show the qualitative associations we have made between the nodes, heat capacities and conductances in the model and the physical parts of the apparatus. Temperatures and heat capacities are associated with the water ($T_w$,$c_w$), the headspace ($T_h$,$c_h$) and the calorimeter airspace ($T_a$,$c_a$) . The temperature of the calorimeter cold plate ($T_c$) has no associated heat capacity because this temperature is treated as a model input, the value of which is not predicted by the model. Thermal conductances are defined between: the water and the headspace ($k_{wh}$); the water and the airspace ($k_{wa}$); the headspace and the airspace ($k_{ha}$); and the airspace and the coldplate ($k_{ac}$). The cubic surface defined by the heat flow sensors forms an energy boundary for the system. Energy flows into the system via electrical leads to the heater and fan that cross this boundary while energy flows out of the system as heat via the effective thermal conductance between the airspace and the coldplate, $k_{ac}$

The differential equations corresponding to the equivalent circuit shown in Fig. 6 are

$$\frac{dT_w}{dt} = \frac{1}{c_w}[Q_{heater} - k_{wa}(T_w - T_a) - k_{wh}(T_w - T_h)] \qquad (13)$$

$$\frac{dT_h}{dt} = \frac{1}{c_h}[k_{wh}(T_w - T_h) - k_{ha}(T_h - T_a)] \qquad (14)$$

$$\frac{dT_a}{dt} = \frac{1}{c_a}[Q_{fan} + k_{wa}(T_w - T_a) + k_{ha}(T_h - T_a) - k_{ac}(T_a - T_c)] \qquad (15)$$

These equations provide the model of the state responses required for step 4 of the protocol. If the model parameters $c_w, c_h, c_a, k_{wa}, k_{wh}, k_{ha}$ and $k_{ac}$ are constants, then equations 13-15 describe a linear heat transfer model.

It is best practice to start the computationally intensive task of parameter estimation with physically reasonable starting values. This is important because the optimization routines for estimating the model parameters can fail if poor initial parameter estimates are provided. Parameter estimates can be obtained by inspection using RC circuit analysis. The temperatures reached during the approximate steady state at 5 W heater input power shortly before $t = 40\ h$ were roughly $T_w = 40\ °C$, $T_h = 30\ °C$, $T_a = 20\ °C$ and $T_c = 18\ °C$. Thus, neglecting the small fan power, $k_{ac} = Q_{heater}/(T_a - T_c) \approx 2.5\ [W/K]$. As the net conductance from $T_w$ to $T_c$ is approximately $k_{wc,net} = Q_{heater}/(T_w - T_c) \approx 0.23\ [W/K]$ and the time-constant at the $T_w$ node is approximately $\tau_w = 1200\ s$, we can very roughly estimate $C_w \approx k_{wc,net}\tau_w = 276\ [J/K]$. Estimates for the other parameters can be obtained with similar calculations.

When mapping real experimental data to an idealized model, it is typically necessary to introduce adjustable offset parameters to account for small offsets in the measured data due to sensor and measurement non-idealities. For this we introduce additional parameters $T_{wo}$, $T_{ho}$ and $T_{ao}$. These offset parameters account for non-ideal temperature offsets in the measured data in the following way:

$$y = x + offsets = \{T_w, T_h, T_a\} + \{T_{wo}, T_{ho}, T_{ao}\} \qquad (16)$$

These offsets give the optimizer additional flexibility to obtain model parameter values which result in predicted outputs $y$ which more closely match the measured temperatures.

For the purposes of computing heat flows from experimental data, the ideal nodal temperatures can be inferred from the experimentally measured temperatures by rearranging equation 16:

$$T_{i,inferred} = T_{i,measured} - T_{io}\ ;\ i = \{w, h, a\} \qquad (17)$$

This is important because the ultimate purpose of our thermal model identification method is to enable the calculation of inferred heat fluxes from the experimentally measured temperatures.

The accuracy with which the three-state model accounts for thermal energy and power can be tested by comparing the inferred heat fluxes with simulated heat fluxes (protocol step 7b). For the three state model, the total input power flowing across the energy boundary shown in Fig. 6 is equal to the power stored within the system (in the three thermal masses $c_w, c_h, c_a$) plus the power flowing back out of the system through the conductance $k_{ac}$. The total input power can be inferred by adding together the inferred stored power and the inferred outflowing power:

$$Q_{in,inferred} = Q_{stored,inferred} + Q_{out,inferred} \qquad (18)$$

where

$$Q_{stored,inferred} = \Sigma_{i=\{w,h,a\}} c_i \frac{d}{dt} T_{i,inferred} \qquad (19)$$

$$Q_{out,inferred} = k_{ac}(T_{a,inferred} - T_{c,inferred}) \qquad (20)$$

When the input power is known, as in the calibration and validation experiments shown here, the stored and outflowing powers can be predicted using the model. The disagreement between these predicted power flows and the inferred power flows is a measure of the power resolution achievable using a given apparatus and model. Here we examine the output power residual

$$e_{Q,out} = Q_{out,inferred} - Q_{out,predicted} \qquad (21)$$

the stored power residual

$$e_{Q,stored} = Q_{stored,inferred} - Q_{stored,predicted} \qquad (22)$$

and the input power residual, which is the sum of the output and stored power residuals:

$$e_{Q_{in}} = e_{Q,out} + e_{Q,stored} \qquad (23)$$

The relative importance of the different residuals depends on the measurement goal. If the goal is to quantify total heat generation by the system under study during a finite time window, $e_{Q,out}$ is the most important residual because eventually all the heat will flow out of the system and be accounted for with an accuracy related to $e_{Q,out}$. If the goal is to measure an instantaneous input power to the system then $e_{Q,in}$ is the most important residual because it describes how accurately the input power is measured. We will see below that the dominant contribution to $e_{Q_{in}}$ comes from the stored power term (Eqn. 19), which, due to the time-derivatives, amplifies processes which cause model inaccuracies on short time scales, such as rapid transients in the input power and high frequency measurement noise.

In the examples below we carry out the calibration and validation aspects of the above protocol for three different models based on the three state example shown above. For each model we estimate the parameters using the calibration data segment (step 5), examine the resulting fit (step 6) and assess the model's power and energy accounting for the calibration data (step 7). Then, we compute the input power residuals for the validation data segment and estimate the calorimetric energy and power measurement accuracy and resolution obtained (step 8).

### *Example 1 - linear three-state model*

This section describes the calibration and validation parts of the protocol (steps 5 thru 8) using the linear three-state model (equations 13-16). This example begins with model parameter estimation (step 5).

Running the estimation using starting parameters obtained by the procedure described above, we find a value for the water thermal mass ($c_w = 403\ JK^{-1}$) that is reasonable: it is within a factor of two of our initial estimate ($276\ JK^{-1}$). The estimate for the airspace thermal mass ($c_a = 774\ JK^{-1}$) is within an order of magnitude of the estimate for $c_w$, which also seems physically reasonable. The headspace thermal mass estimate ($c_h = -198\ JK^{-1}$), however, is negative and therefore non-physical. Similarly non-physical negative parameter estimates were obtained for the conductances connected to the headspace heat capacity, $k_{wh0}$ and $k_{ha}$. These non-physical parameter estimates provide a strong indication that the linear three-state model does not provide an adequate description of the thermal system.

The three state model also makes use of small temperature offsets (~0.5 K or smaller) which are plausible given that the RTD thermometers employed in this study were used as received without calibration. We only needed to correct for thermometry offset errors because small errors in the scaling of resistance thermometer resistances into temperatures can also be absorbed into the adjustable parameters of the model. All the parameter estimates for the linear three state model (and the other models) are shown in Table 2.

**Table 2** Estimated fit parameters and uncertainties obtained by fitting the models to the calibration data segment. The parameter uncertainties shown are standard deviations returned by the MATLAB® *nlgreyest* command, which extracts them from diagonal of the fit covariance matrix. The uncertainties on the conductances and heat capacities decrease significantly when the three state model is improved by introducing non-linearity into the $k_{wh}$ conductance. Non-physical negative parameter estimates for $c_h, k_{wh}$ and $k_{ha}$ are returned by the three state linear model but not by the other two models

| parameter | value ± standard deviation | | |
|---|---|---|---|
| | 3 state linear model | 3 state non-linear model | Hybrid 3 state model |
| $c_w$ | 403.3 ± 8.0 JK$^{-1}$ | 318.07 ± 0.19 JK$^{-1}$ | 342.87 ± 0.14 JK$^{-1}$ |
| $c_h$ | -198 ± 20 JK$^{-1}$ | 24.11 ± 0.29 JK$^{-1}$ | 0.8365 ± 0.0089 JK$^{-1}$ |
| $c_a$ | 774 ± 49 JK$^{-1}$ | 190.6 ± 2.0 JK$^{-1}$ | 319.26 ± 0.96 JK$^{-1}$ |
| $k_{wa}$ | 1.33 ± 0.14 WK$^{-1}$ | 0.14459 ± 0.00062 WK$^{-1}$ | 0.178393 ± 0.00094 WK$^{-1}$ |
| $k_{wh}$ or $k_{wh0}$ | -2.64 ± 0.36 WK$^{-1}$ | 0.3198 ± 0.0066 WK$^{-1}$ | 0.1390 ± 0.0077 WK$^{-1}$ |
| $k_{wh1}$ | n/a | -0.01063 ± 0.00041 WK$^{-2}$ | -0.00107 ± 0.00047 WK$^{-2}$ |
| $k_{wh2}$ | n/a | 309.3E-6 ± 6.6E-6 WK$^{-3}$ | 112.7E-6 ± 7.5E-06 WK$^{-3}$ |
| $k_{ha}$ | -1.75 ± 0.23 WK$^{-1}$ | 0.2222 ± 0.0010 WK$^{-1}$ | 0.1633 ± 0.0016 WK$^{-1}$ |
| $k_{ac}$ | 2.55173 ± 0.00053 WK$^{-1}$ | 2.55197 ± 0.00040 WK$^{-1}$ | 2.52900 ± 0.00018 WK$^{-1}$ |
| $T_{wo}$ | 0.5301 ± 0.0026 K | 0.4265 ± 0.0013 K | 0.4195 ± 0.0022 K |
| $T_{ho}$ | 0.3297 ± 0.0020 K | 0.38778 ± 0.00098 K | 0.3813 ± 0.0015 K |
| $T_{ao}$ | -0.07253 ± 0.00016 K | -0.07243 ± 0.00012 K | n/a |
| $V_{s0}$ | n/a | n/a | 5.925 ± 0.014 mV |
| $V_{s1}$ | n/a | n/a | 274 mV K$^{-1}$(fixed) |

The MATLAB system ID toolbox reports the fit quality using the normalized root mean square error (NRMSE) which is defined in equation S1 of supplementary material (online resource 1). A related metric which reports on the absolute error is the root mean square (equation S2, online resource 1). Examining the model fits, NRMSE values and fit temperature residuals shown in Figs. 7a and 7d provides additional information about the inadequacies of the linear three state model. In particular, the residuals contain two clear features:

1. **Sharp spikes near large changes in the input power.** These spikes arise due to the coarseness of the lumped element thermal model. While in reality temperature changes due to changes in the input power must propagate through the spatially distributed thermal system between the heater and the temperature sensors, in the model changes in input power affect the temperature sensors instantaneously. This discrepancy between the model and the real system results in large transient model errors when the input power changes rapidly.

2. **Negatively correlated values of the $T_w$ and $T_h$ residuals which change amplitude and sign as the input power increases.** This residual structure suggests that these nodes are coupled in a way not captured by the linear model. The optimizer found a "best compromise" set of parameters which minimized the model output prediction error (equation 4) by underpredicting $T_h$ at low temperature and over predicting it at high temperature (and vice-versa for $T_w$). Systematic temperature dependencies such as these in the residuals suggest that a non-linear element may be necessary to describe the thermal system. In the thermal system studied here, non-linear heat transport from the water to the headspace due to evaporation and condensation is likely occurring. In the next example, we will see that introducing an appropriate non-linear conductance between the $T_w$ and $T_h$ nodes largely eliminates this negatively correlated structure of the $T_w$ and $T_h$ residuals (Fig. 7e), while leaving the sharp spike features essentially unaffected.

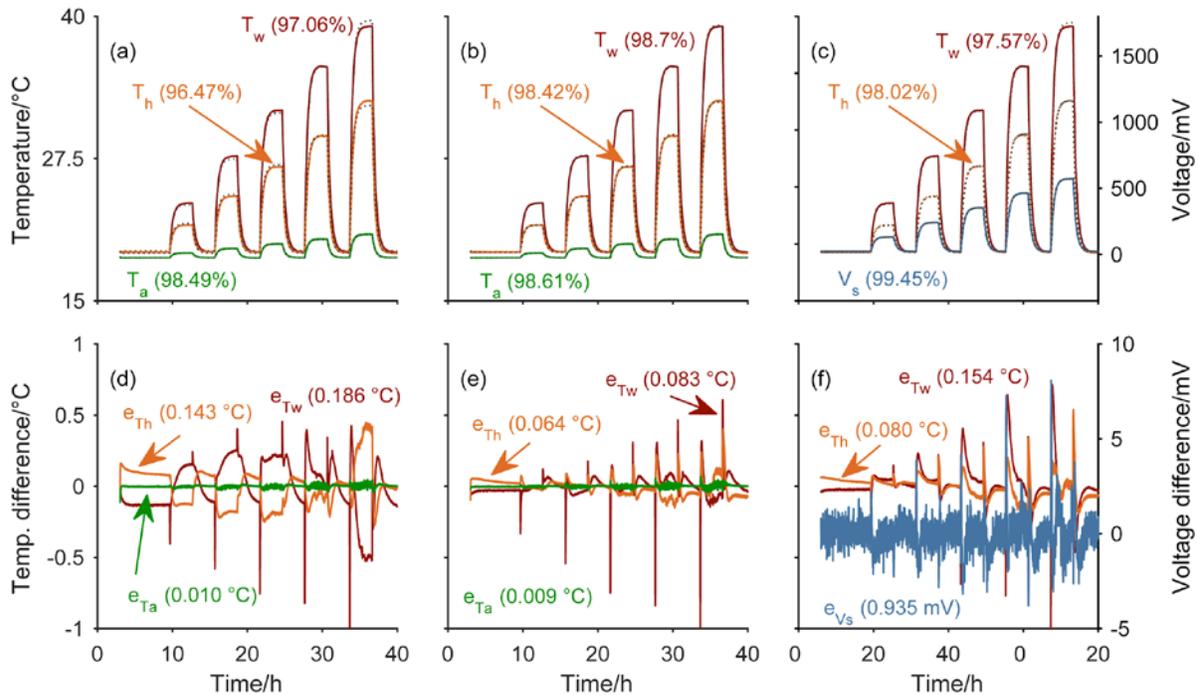

**Fig. 7** Fits for each model to the calibration data segment. The data (solid lines), model fits (dashed lines) and NRMSE values for each model output are shown in **a** for the three state linear model, in **b** for the three state non-linear model and in **c** for the hybrid model. The fit residuals (measured less predicted outputs) and RMS values for each model output are shown in **d** for the three state linear model, in **e** for the three state non-linear model and in **f** for the hybrid model

The inferred input power and input power residuals for the calibration data (step 7) are shown in Figs. 8a and 8b. The sharp spikes observed in the temperature residuals are also present in the power residuals, and for the same reason. Unlike the temperature residuals, the inferred power also contains a significant amount of noise. This noise arises largely due to measurement noise on the $T_a$ RTD, which is aggravated by the numerical differentiation used in our implementation of equation 21.

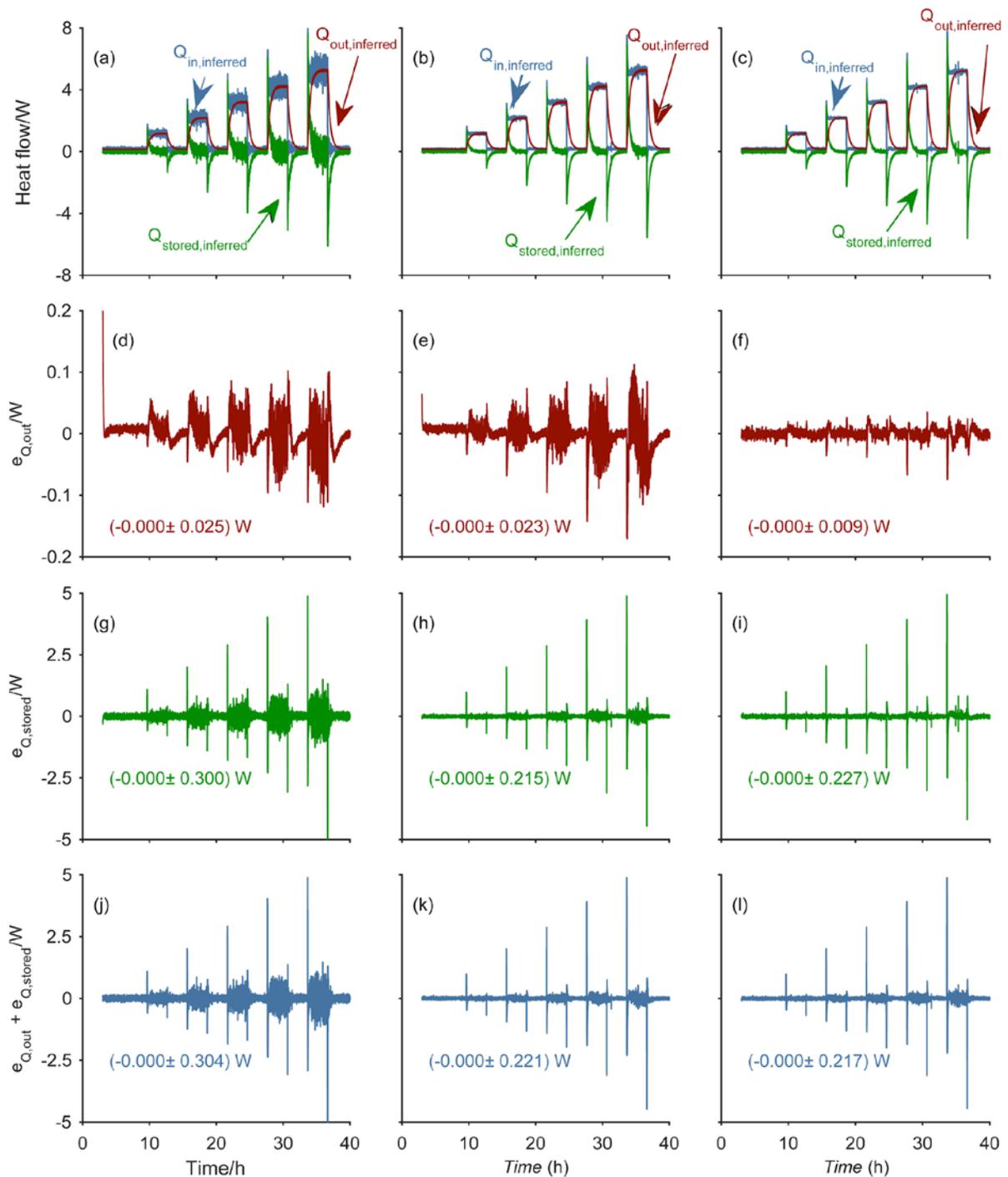

**Fig. 8** The inferred output power, inferred stored power and inferred input power for the calibration data segment are shown in **a** for the linear three state model, in **b** for the non-linear three state model and in **c** for the hybrid three state model. The residual output powers for each model are shown in **d**,**e** and **f**. The residual stored powers are shown in **g**,**h** and **i**. The sum of the stored and output power residuals, which is clearly dominated by the stored power residual, is shown in **j**,**k** and **l**. The mean ± standard deviation of the power residuals are indicated

Once the three state linear model parameters were estimated, this was validated by predicting the experimentally measured temperatures (step 8a) for the validation segment based only on the model and the inputs (Figs. 9a and 9d). Tests of predictive accuracy are crucial for proving two important things:

- that the apparatus is stable in time
- that the model generalizes to other data sets and is not overfitting the calibration data

We found that the prediction accuracy was comparable to the fit accuracy, indicating that the model can fit data other than the calibration data and that the apparatus is relatively stable over the 80 h time span of the experiment. The major features in the temperature prediction residuals (spikes and negatively correlated $T_w$ and $T_h$ residuals) are similar to those for the calibration segment, indicating that the three-state linear model's inadequacies are also repeatable. Although the linear three state model appears to make valid power estimates despite it's non-physical parameter values, measurements based on non-physical models should be avoided because such models may fail in nonsensical ways under conditions different than the calibration conditions. For example, if, in the validation segment of the present experiment, a heat input appeared in the heat-source headspace, the experimentally measured temperature $T_a$ would increase while the model would predict a temperature decrease.

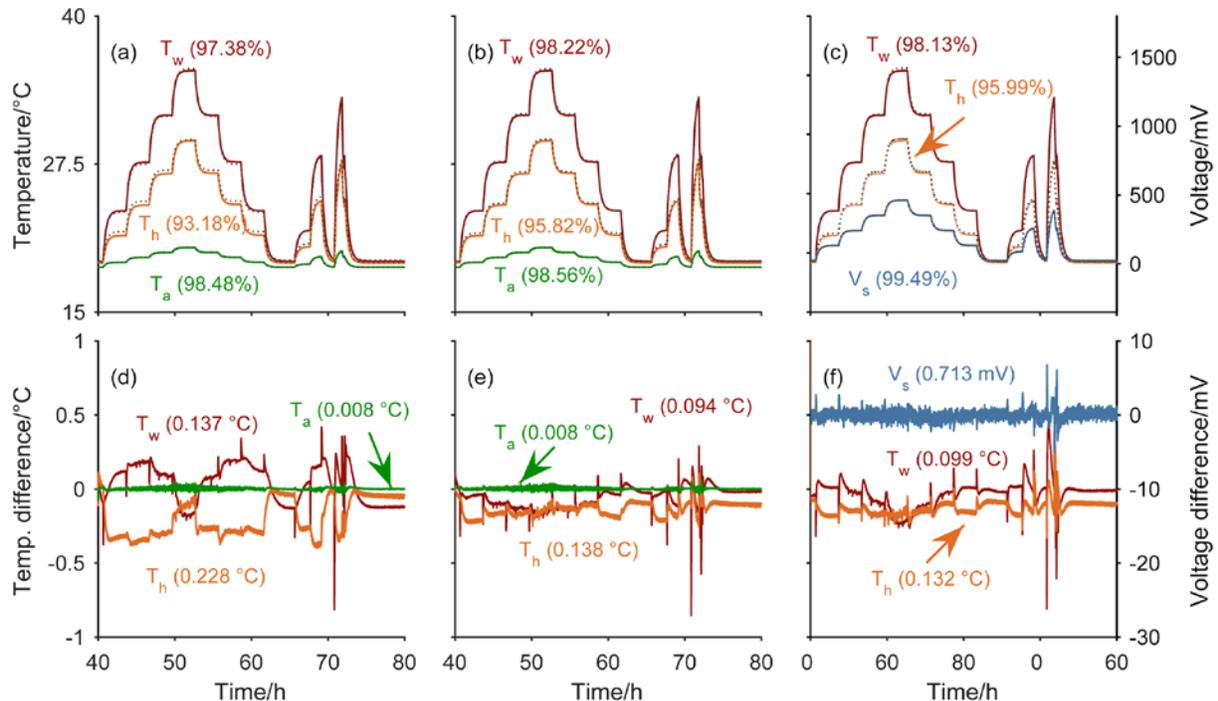

**Fig. 9** Output predictions made by each model for the validation data segment. The measured outputs (solid lines), predicted outputs (dashed lines) and NRMSE are shown in **a** for the linear three state model, in **b** for the non-linear three state model and in **c** for the four state model. The prediction residuals and their RMS values for the linear three state, non-linear three state and hybrid three state models are shown in **d**, **e** and **f** respectively

Further model validation (step 8b) involved using the measured sensor temperatures from the validation data segment to infer the stored, outflowing and input powers as per Eqns. 18-20. This estimate was then compared to the known input power (Figs. 10a and 9d). The features in the inferred power residual for the validation segment are similar to those observed for the calibration segment.

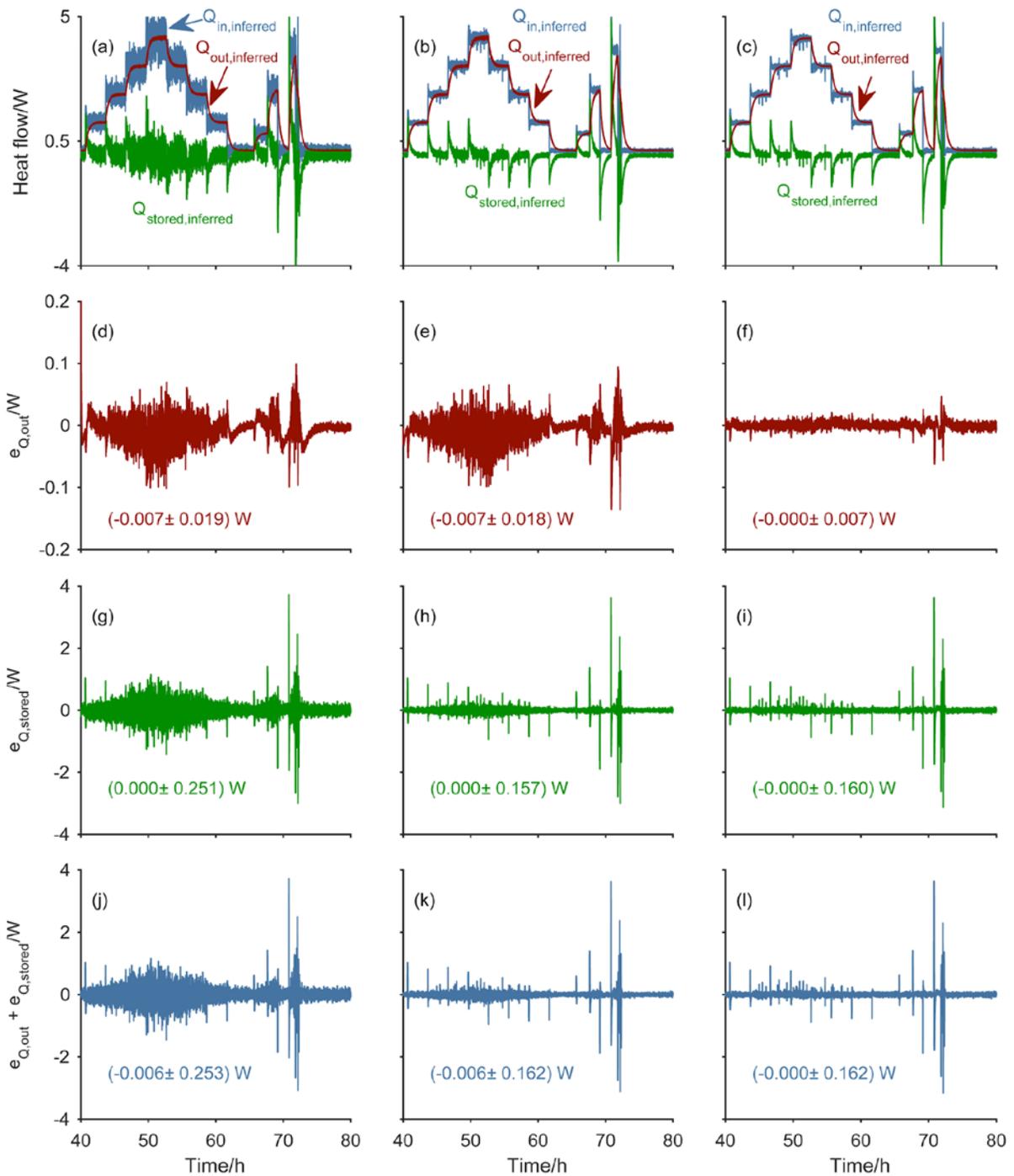

**Fig. 10** The inferred output power, inferred stored power and inferred input power for the validation data segment are shown in **a** for the linear three state model, in **b** for the non-linear three state model and in **c** for the hybrid three state model. The residual output powers for each model are shown in **d**,**e** and **f**. The residual stored powers are shown in **g**,**h** and **i**. The sum of the stored and output power residuals is shown in **j**,**k** and **l**. The mean ± standard deviation of the power residuals are indicated

**Table 3** Power and energy analysis for each model applied to the validation data. The residual power $Q_{in,residual}$ in the first row is described in terms of its mean, root-mean-square (RMS), minimum and maximum values both in absolute units and as percentages of the 1.594 W mean input power during the validation segment

| quantity | 3 state linear model | 3 state non-linear model | Hybrid 3 state model |
|---|---|---|---|
| $e_{Q,out} = Q_{out,inferred}(t) - Q_{out,modelled}(t)$ | **Mean:** -6.62 mW (-0.42%) **RMS:** 19.33 mW (1.21%) **Max.:** 206.15 mW (12.93%) **Min.:** -101.26mW (-6.35%) | **Mean:** -6.66 mW (-0.42%) **RMS:** 18.12 mW (1.14%) **Max.:** 94.14 mW (5.91%) **Min.:** -135.85 mW (8.52%) | **Mean:** -0.19 mW (-0.01%) **RMS:** 6.57 mW (0.41 %) **Max.:** 46.74 mW (2.93%) **Min.:** -62.14 mW (-3.90%) |
| $e_{Q,stored} = Q_{stored,inferred}(t) - Q_{stored,modelled}(t)$ | **Mean:** 0.13 mW (0.01%) **RMS:** 251.05 mW (15.75%) **Max.:** 3723.02 mW (233.57%) **Min.:** -3002.59 mW (-188.37%) | **Mean:** 0.31 mW (0.02%) **RMS:** 157.16 mW (9.86%) **Max.:** 3631.15 mW (227.81%) **Min.:** -3005.59 mW (-188.56 %) | **Mean:** -0.10 mW (-0.01%) **RMS:** 159.58 mW (10.01%) **Max.:** 3635.04 mW (228.05%) **Min.:** -3117.56 mW (195.59%) |
| $e_{Q,in} = e_{Q,out} + e_{Q,stored}$ | **Mean:** -6.49 mW (-0.41%) **RMS:** 253.49 mW (15.90%) **Max.:** 3714.43 mW (233.03%) **Min.:** -3073.93 mW (192.85%) | **Mean:** -6.35 mW (-0.40%) **RMS:** 161.86 mW (10.15%) **Max.:** 3627.75 mW (227.59%) **Min.:** -3110.92 mW (-195.17%) | **Mean:** 0.11 mW (0.01%) **RMS:** 159.20 mW (9.99%) **Max.:** 3635.44 mW (228.08%) **Min.:** -3057.38 mW (-191.81%) |
| $E_{in,measured} = \int_{t=40\,h}^{t=80\,h} Q_{in,measured}(t')dt'$ | 229.543 kJ | 229.543 kJ | 229.543 kJ |
| $E_{in,inferred} = \int_{t=40\,h}^{t=80\,h} Q_{in,inferred}(t')dt'$ | 230.480 kJ | 230.458 kJ | 229.586 kJ |
| $E_{in,inferred} - E_{in,measured}$ | 0.936 kJ | 0.914 kJ | 0.042 kJ |
| $\dfrac{E_{in,inferred} - E_{in,measured}}{E_{in,measured}}$ | 0.41% | 0.40% | 0.02% |

To summarize example 1, we found that the linear three-state model was able to predict the sensor temperatures to within 0.25 °C for temperature excursions of amplitude ~25 °C. Performing power and energy analysis on the model results (Table 3) illustrates that the accuracy in the the total input energy measurement (~0.4%) is much better than the worst case instantaneous power accuracy (up to 233% of the average input power). The worst instantaneous power errors occurred briefly during rapid input power changes. Except near these rapid changes, the input power during the validation segment could be inferred with moderate accuracy: the standard deviation of the input power residual was 15.9% of the mean input power. The principal shortcomings of the model were a failure to deal with rapid changes in input power and failure to accurately fit the outputs at all inputs powers across the range of input powers employed, likely due to an unmodelled non-linearity in the system.

## Example 2 - non-linear three-state model

In this example, we address the failure of the linear three state model to accurately fit the temperature data across the full range of input powers by adding a non-linear conductance to the model.

We implemented a non-linear version of the three-state model by allowing the conductance $k_{wh}$ to vary as a function of temperature in the following way

$$k_{wh} = k_{wh0} + k_{wh1}T_w + k_{wh2}T_w^2 \qquad (24)$$

This non-linear conductance between the water ($T_w$) and headspace ($T_h$) nodes was added to capture some of the expected non-linear heat transport that may be occuring in the heat source due to evaporation and condensation. The addition of this non-linear conductance decreased the fit residuals at the $T_w$ and $T_h$ nodes by roughly a factor of two relative to the linear model (Figs. 7b,e). The negatively correlated behaviour disappeared and the temperature dependence of the residual decreased. Worth noting is that the introduction of the non-linearity into the $k_{wh}$ conductance (through the non-linear parameters $k_{wh1}$ and $k_{wh2}$) causes changes in many of the other model parameters (see Table 2). In particular, non-physical negative parameter estimates no longer occur. The non-linear conductance also improves the temperature-prediction accuracy for the $T_w$ and $T_h$ nodes (Fig. 9b,e) by a factor of more than 1.5.

For both the validation and the calibration segment, the non-linear three-state model exhibits roughly 30% smaller stored power residuals than the linear model (compare Figs. 8g,h with Figs. 10g,h). This is because temperature measurement noise at the $T_w$, $T_h$ and $T_a$ nodes contributes noise to the stored power in proportion to the magnitude of the $c_w$, $c_h$ and $c_a$ parameters and these thermal mass parameters are significantly smaller for the non-linear model than for the linear model. The extremal values of the stored power residuals - which occur due to model inaccuracy at rapid input power transitions - are about 3.7 W for both the linear and non-linear models (Table 3). This shows that the introduction of the nonlinearity does not improve the ability of the model to capture the system response to high-frequency perturbations; for that, larger lumped element models with topologies guided by finite element simulations would be helpful [25].

The output power residuals (figs 8e,10e) were improved only very slightly (1-2 mW decrease in standard deviation) by the addition of the non-linear conductance. This is consistent with the small improvement to the fit at the $T_a$ node (compare figs 7a,b). The major contributions to the output power residuals are the spikes due to rapid input power changes and measurement noise on the $T_a$ RTD, neither of which are addressed by the added non-linearity. In the third and final example, this measurement noise is addressed by replacing the noisy $T_a$ temperature measurement with a low-noise Seebeck-effect heat flow measurement.

## Example 3 - hybrid three-state model incorporating both temperature and heat flow measurements

The models described up to this point can be further extended to work not just with multiple sensors but with multiple *types* of sensors. This example describes a hybrid model incorporating both temperature and heat flow measurements. Specifically, this model uses the Seebeck voltage $V_s$ from the calorimeter heat-flow sensor array as an output instead of the noisier $T_a$ temperature measurement.

Unlike in the previous models where all the nodal temperatures were directly associated with a temperature measurement, in this version of the three-state model $T_a$ is a hidden state which has no directly associated measurement.  The Seebeck voltage $V_s$ produced by the calorimeter is, however, related to $T_a$ in this model as follows:

$$V_s = V_{so} + V_{s1}(T_a - T_c) \qquad (25)$$

where $V_{so}$ and $V_{s1}$ are new model parameters for the heat flow sensor array voltage offset and differential temperature response, respectively.  If estimates of these parameters are available, Eqn. 24 can be rearranged to obtain an inferred value of $T_a$ from the experimentally measured value of $V_{seebeck}$.  It is important to keep in mind that because of the extremely coarse nature of the lumped element models used here, the "locations" associated with the temperature nodes are highly qualitative and depend strongly on the associated sensors.  For this reason, the unmeasured $T_a$ state used in this example (see Eqn. 25) is potentially very different than the measured airspace temperature used in the previous examples and should be considered unrelated despite the fact that the model topology shown in Fig. 6 still applies (except for the now-invalid association of the $T_a$ node with the airspace RTD).  For this reason, we do not attempt to simultaneously fit the model to the seebeck voltage and the $T_a$ temperature measurement.

The state equations for this model are the same as the nonlinear three state model above, but the outputs are now:

$$y = \{T_w + T_{wo}, T_h + T_{ho}, V_s\} \qquad (26)$$

For the hybrid model, the stored power is inferred as before (Eqn. 19) but the power conducted out of the system is now inferred from $V_s$ as

$$Q_{out,inferred} = \frac{k_{ac}(V_S - V_{so})}{V_{s1}} \qquad (27)$$

The model parameters obtained by fitting this hybrid model to the calibration data segment are reported in Table 2 and the fit and residuals are shown in Figs. 7c and 7f. Because the heat flow $Q_{ac}$ depends only on the ratio of $k_{ac}$ and $V_{s1}$, $V_{s1}$ was fixed at 274 mVK$^{-1}$ in order to properly constrain the optimization problem; this value results in $k_{ac}$ values that give physically reasonable temperature excursions at $T_a$ of a few degrees celsius.  The NRMSE for the hybrid model fit to $V_s$ (99.45%,) was better than the best previous fit to $T_a$ (98.61%) (compare Figs. 7b,c).   Figs. 9c and f show that, in the validation segment, the hybrid model predicts the water and headspace temperatures with similar accuracy to the previous model, but that the seebeck voltage is again predicted more accurately on a NRMSE basis than the airspace temperature measurement which it replaced.  This predictive improvement translates into an improved accuracy in the inferred power, as evidenced by the greater than two-fold decrease in the standard deviation of the output power residuals for both the calibration and validation segments to below 10 mW in both cases (compare Figs. 8e,f with Figs. 10e,f).  The mean output power residual also improves significantly (from -6.66 mW to -0.19 mW, Table 3), yielding a corresponding order of magnitude improvement to the accuracy of the energy accounting (0.02% error vs ≥0.40% for the other models, Table 3).

The reduced standard deviation of the output power residual is due to the reduced measurement noise on the Seebeck heat-flow sensor array relative to the airspace RTD which it replaced.  It seems unlikely, however, that this reduction in zero-mean electrical noise explains the reduction in the mean output power residual.  That improvement is likely due to the spatially distributed nature of the Seebeck heat flow sensor array: a spatially averaged measurement is more appropriate than a point measurement for use with low-dimensional lumped element models, because these models deal intrinsically with spatially averaged quantities.

One advantage of the lumped-element grey box models used here is that a number of different heat flows can be calculated. Up to this point, the total power stored in the the system and the total power conducted out of the system have been calculated however other heat flows may also be of interest. For example, we may wish measure the heat flowing out of the heat-source (see Fig. 11). Using the sign conventions in Fig. 11, energy balance at node T$_a$ gives:

$$Q_{ha} + Q_{wa} + Q_{fan} = Q_{a,stored} + Q_{ac} = c_a \frac{dT_a}{dt} + k_{ac}(T_a - T_c) \quad (28)$$

$$Q_{out,from\ heat\ source} = Q_{ha} + Q_{wa} = Q_{a,stored} + Q_{ac} - Q_{fan} = c_a \frac{dT_a}{dt} + k_{ac}(T_a - T_c) - Q_{fan} \quad (29)$$

In the three state hybrid model, $T_a$ can be inferred from the measured Seebeck voltage $V_s$ and the measured temperature $T_c$ by inverting equation 25 as follows:

$$T_{a,inferred} = \frac{1}{V_{s1}}(V_{s,measured} - V_{so}) + T_{c,measured} \quad (30)$$

If $Q_{fan}$ is also known known (it is directly measured by the power supply in this case), we can then infer $Q_{out,from\ heat\ source}$ using $T_{a,inferred}$ and $Q_{fan}$ according to equation 29. As seen in Fig. 11a, the heat flow out of the heat source changes more slowly than the power input into the heat source. As seen in Fig. 11b, this more slowly changing heat flow can be recovered with a much smaller root-mean-square error than can the input power and the amplitude of the large spikes in the residual power is also significantly reduced.

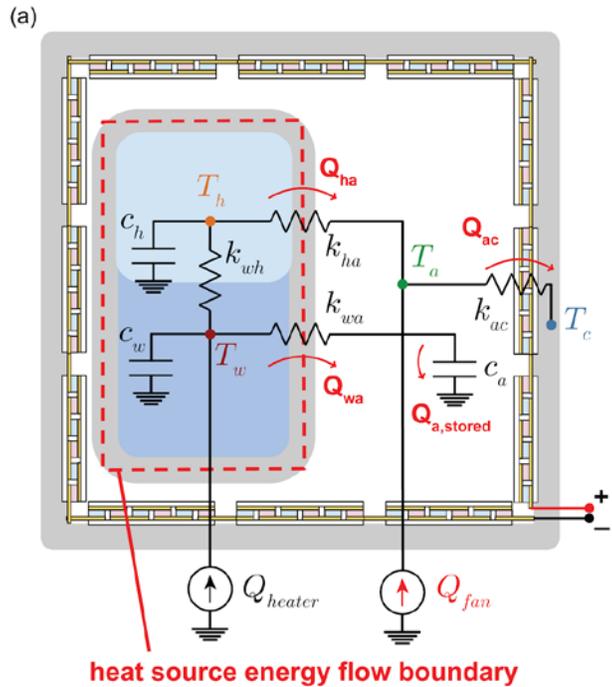
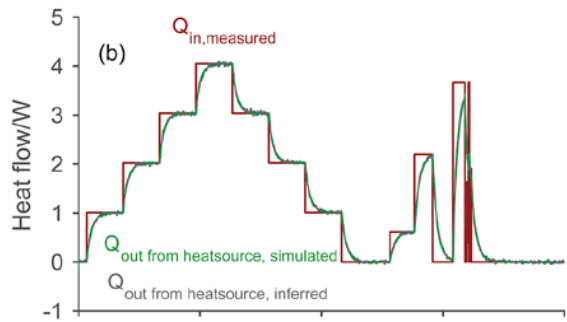
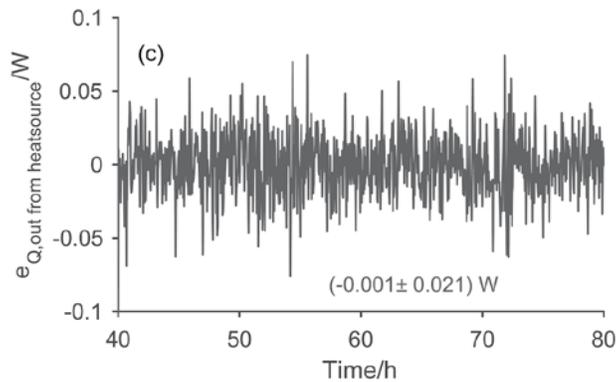

**Fig. 11** Study of the power leaving the heat source. **a** diagram of the relevant energy boundary and heat flows **b** measured input power into the heat source and modelled and inferred heat flows leaving the heat source **c** residual of the heat flow out of the heat source (inferred less predicted). The mean ± standard deviation of the power residual is indicated

# Conclusions

We have shown that system identification calorimetry is a versatile tool suitable for analyzing heat flows in thermal systems. In particular, this method is useful for modelling systems with non-ideal geometries, where non-linear thermal conductances or heat capacities are important or where multiple thermal masses are relevant to the system dynamics.

The three-state model used here was developed by examining the geometry of the apparatus and applying physical intuition to determine an appropriate model which was then fit to calibration data. This model was subsequently improved by examining the temperature and power residuals and making targeted improvements based on features observed in these residuals. First a non-linear thermal conductance was added to the model, which improved temperature predictions. Then a distributed heat flow sensor was used to replace a single-point thermometer measurement, demonstrating the fusion of different types of thermal sensors in a single model and reducing the noise and increasing the accuracy of the inferred heat flux.

In the thermal system studied here, our system identification calorimetry technique enabled the measurement of the total input energy during a validation experiment with an accuracy at the 0.02% level. The instantaneous power flowing out of the calorimeter was measured with a RMS error of 0.41% of the mean input power and a worst-case instantaneous error of 3.90% of the input power. The more rapidly changing input power was less accurately measured: the RMS input power residual was 10% of the mean input power. The rapid changes in the input power also caused much larger transient errors in the inferred input and stored powers (up to 228% of the mean input power). These large transient errors are expected because rapid processes are not accurately captured by low order lumped element models [26]. Reduced transient errors are expected for less rapidly changing heat flows, or for higher order models which more accurately approximate the true continuum thermal system. Such a reduction was observed when a more slowly changing heat flow was examined (Fig. 11).

The measurement of heat flows is important to a large number of scientific and engineering fields. While for didactic purposes this tutorial uses data from a relatively well-behaved model thermal system, we have previously demonstrated the use of system identification calorimetry in less ideal circumstances [8]. We contend that system identification calorimetry has value for estimating heat flows in a variety of applications where conventional calorimetric measurements are infeasible. The development of tools to facilitate the use of this method could be a valuable direction for future work.

# Notation

$Q$ - rate of heat flow per unit time [Watts]
$e_x$ - residual of quantity X (i.e. $X_{measured}$ - $X_{predicted}$) [units of X]
$c$ - integrated heat capacity (thermal mass) [J/K]
$k$ - net thermal conductance [W/K]
$T$ - temperature [K or °C]
$\boldsymbol{\theta}$ - vector of model parameters
$\boldsymbol{x}$ - vector of model states
$\boldsymbol{u}$ - vector of model inputs
$\boldsymbol{y}$ - vector of model outputs

# Acknowledgements

We acknowledge Google LLC for financial support and Dr. Edmund Storms for his detailed input on the Seebeck calorimeter design.  This research was supported in part from the Canada First Research Excellence Fund, Quantum Materials and Future Technologies Program, Natural Sciences and Engineering Research Council of Canada, Canada Research Chairs, and the Stewart Blusson Quantum Matter Institute (SBQMI). BPM acknowledges support from the SBQMI's Quantum Electronic Science and Technology Initiative. We thank the SBQMI machine shop for assistance with instrument fabrication.

# References

1. Deng K, Barooah P, Mehta PG, Meyn SP. Building Thermal Model Reduction via Aggregation of States. Am Control Conf. 2010. p. 5118–23.
2. Banerjee S, Cole JV, Jensen KF. Nonlinear model reduction strategies for rapid thermal processing systems. IEEE Trans Semicond Manuf. 1998;11:266–75.
3. Zielenkiewicz W. Calorimetric models. J Therm Anal. 1988;33:7–13.
4. O'Neill MJ. The Analysis of a Temperature-Controlled Scanning Calorimeter. Anal Chem. 1964;36:1238–45.
5. Höhne GWH, Hemminger WF, Flammersheim H-J. Differential Scanning Calorimetry. 2nd ed. Berlin Heidelberg: Springer-Verlag; 2003.
6. Zielenkiewicz W. Thermal-dynamic analogy method in calorimetry. J Therm Anal Calorim. 2007;88:59–63.
7. Drebushchak VA. From electrical analog to thermophysical modeling of DSC. J Therm Anal Calorim. 2011;105:495–500.
8. Macleod BP, Schauer PA, Hu K, Lam B, Fork DK. High-temperature high-pressure calorimeter for studying gram- scale heterogeneous chemical reactions.  Rev. Sci. Instrum. 2017;88:084101-1-8
9. Ljung L. System identification: Theory for the user. 2nd ed. Upper Saddle River, NJ: Prentice Hall; 1999.
10. Bohlin TP. Practical Grey-box Process Identification: Theory and Applications. London, England: Springer-Verlag; 2006.
11. Point R, Petit JL, Gravelle PC. Reconstruction of thermokinetics from calorimetric data by mean of numerical inverse filters. 1979;17:383–93.
12. Macqueron JL, Ortin J, Thomas G, Torra V. Thermogenesis: Comparative efficiency of deconvolution based on optimal control and inverse filters. Thermochim Acta. 1983;67:213–22.
13. Ortín J, Ramos A, Torra V. Thermogenesis: An approach to nearly exact deconvolution in time-varying systems. Thermochim Acta. 1985;84:255–62.
14. Socorro F, Rodríguez De Rivera M, Jesus C. A thermal model of a flow calorimeter. J Therm Anal Calorim. 2001;64:357–66.
15. Schetelat P, Etay J. Inductive modulated calorimetry: analytical model versus numerical simulation. COMPEL. 2008;27:436–44.
16. Schetelat P, Etay J. A new approach for non-contact calorimetry: System identification using pseudo-white noise perturbation. Heat Mass Transf. 2011;47:759–69.
17. Schetelat P. Modélisation et simulation de la calorimétrie modulée inductive. Institut National Polytechnique de Grenoble; 2010.
18. Sohlberg B. Grey box modelling for model predictive control of a heating process. J Process Control. 2003;13:225–38.
19. Akkari E, Chevallier S, Boillereaux L. A 2D non-linear "grey-box" model dedicated to microwave thawing: Theoretical and experimental investigation. Comput Chem Eng. 2005;30:321–8.
20. George Sidebotham. Heat Transfer Modeling: an inductive approach. Cham, Switzerland: Springer International Publishing; 2015.
21. Nise N. Control systems engineering. 6th ed. Hoboken, NJ: Wiley and Sons; 2011.
22. El-Nasr AA, El-Haggar SM. Effective thermal conductivity of heat pipes. Heat Mass Transf. 1996;32:97–101.
23. Benzinger TH, Kitzinger C. Direct calorimetry by means of the gradient principle. Rev Sci Instrum.

1949;20:849–60.
24. Meis SJ, Dove EL, Bell EF, Thompson CM, Glatzl-Hawlik MA, Gants AL, et al. A gradient-layer calorimeter for measurement of energy expenditure of infants. Am J Physiol. 1994;266:R1052-60.
25. Christiaens F, Vandevelde B, Beyne E, Mertens R, Berghmans J. A generic methodology for deriving compact dynamic thermal models, applied to the PSGA package. IEEE Trans Components, Packag Manuf Technol Part A. 21:565–76.
26. Barbaro S, Giaconia C, Orioli A. Analysis of the accuracy in modelling of transient heat conduction in plane slabs. Build Environ. 1986;21:81–7

# Electronic supplementary material for *System identification calorimetry* – online resource 1

B. P. MacLeod*,[1,4] (ORCID: 0000-0002-8547-9318)

D. K. Fork*,[2] (ORCID: 0000-0001-9559-1277)

B. Lam[1]

C. P. Berlinguette*,[1,3,4,a] (ORCID: 0000-0001-6875-849X)

[1]*Department of Chemistry, University of British Columbia, Vancouver, V6T1Z1, Canada*
[2]*Google LLC, 1600 Amphitheatre Pkwy, Mountain View, California, 94043, USA*
[3]*Department of Chemical and Biological Engineering, University of British Columbia, Vancouver, V6T 1Z3, Canada*
[4]*Stewart Blusson Quantum Matter Institute, University of British Columbia,  Vancouver, V6T 1Z4, Canada*
a) Author to whom correspondence should be addressed. Electronic mail: cberling@chem.ubc.ca

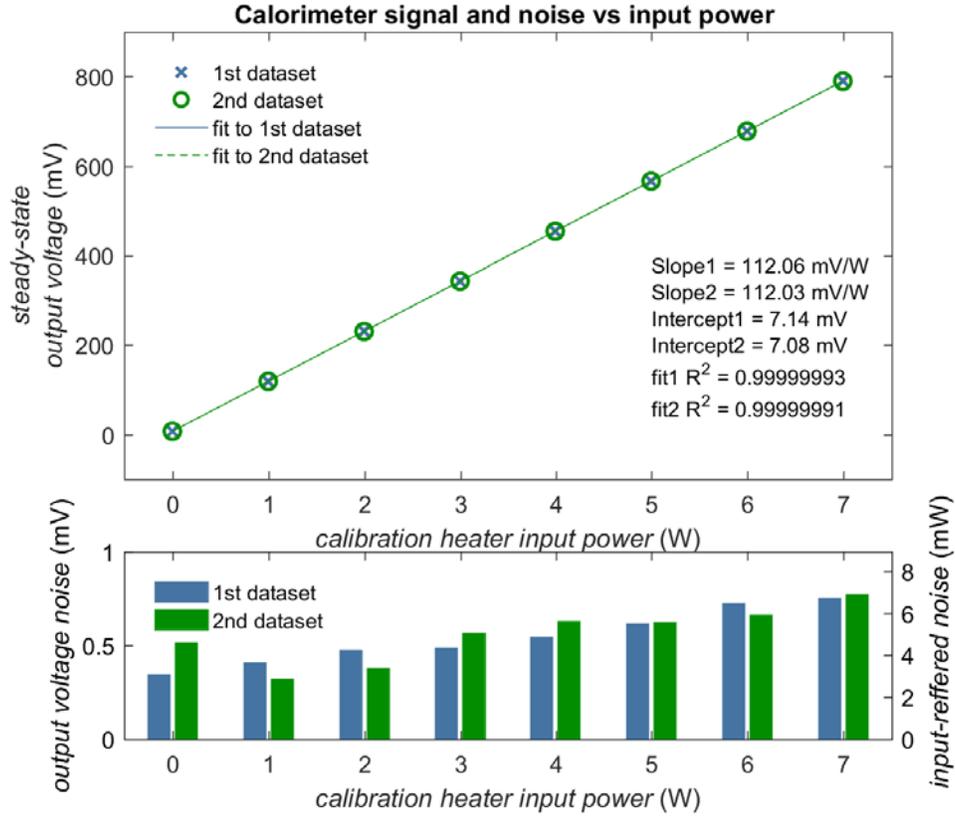

**Fig. S1:** Assessment of the signal and noise response of the calorimeter to steady-state heating. Two datasets each about 50 h long were acquired. In each data set the calorimeter was allowed 3 h to come to steady-state at heater input powers of 0,1,2,3,4,5,6 and 7 W. The steady state signal (top panel) was obtained at each input power. The output voltage noise at each input power level (bottom panel) was characterized by taking the standard deviation of the voltage over a 30-minute window during which the calorimeter was approximately at thermal equilibrium at each input power level. Based on this data a steady-state power measurement resolution, computed here as three times the standard deviation of the equilibrium signal, of 21 mW or less for input powers up to 7 W is obtained.

In the MATLAB system ID toolbox, the agreement between the model outputs and the associated measured data is reported as a normalized root mean square error (NRMSE) which is defined for the i-th output as

$$NRMSE_i = 1 - \frac{[\sum_{t=1}^{N_{samples}}(y_{i,measured}(t) - y_{i,simulated}(t))^2]^{1/2}}{[\sum_{t=1}^{N_{samples}}(y_{i,measured}(t) - \overline{y_{i,measured}})^2]^{1/2}} \tag{S1}$$

A related quantity which describes the fit quality in an absolute sense is the root mean square (RMS) of the fit residual which is defined for the i-th output as

$$RMS = [\frac{1}{N_{samples}}\sum_{t=1}^{N_{samples}}(y_{i,measured}(t) - y_{i,simulated}(t))^2]^{1/2} \tag{S2}$$